\date{}
\documentclass[preprint]{revtex4}
\usepackage{graphicx}
\begin{document}

\title {Casimir interaction between two concentric cylinders: exact versus
semiclassical results}

\author{
Francisco D.\ Mazzitelli$^{1}$\footnote[2]{Member of CONICET, Argentina.}
\thanks{fmazzi@df.uba.ar}, Mar\'\i a J.\ S\'anchez$^{1 \dag}$
\thanks{majo@df.uba.ar},  Norberto N.\ Scoccola$^{2,3 \dag}$
\thanks{scoccola@tandar.cnea.gov.ar}
and Javier von Stecher$^{1}$ \thanks{von-stecher@ciudad.com.ar} }

\affiliation{
$^1$ Departamento de F\'\i sica {\it J.J. Giambiagi}, Facultad de Ciencias Exactas y Naturales,
Universidad de Buenos Aires - Ciudad Universitaria, Pabell\' on I,
(1428) Buenos Aires, Argentina. \\
$^2$ Departmento de F\'\i sica, Comisi\'on Nacional de
Energ\'\i a At\'omica\\Av. Libertador 8250, (1429) Buenos Aires, Argentina. \\
$^3$ Universidad Favaloro, Sol{\'\i}s 453, (1078) Buenos Aires, Argentina.}

\begin{abstract}

The Casimir interaction between  two perfectly conducting, infinite,
concentric cylinders is computed using a
semiclassical approximation that takes into account families of classical
periodic orbits that reflect off both cylinders. It is then compared
with the exact result obtained by the mode-by-mode summation technique.
We analyze the validity
of the semiclassical approximation and show that it improves the
results obtained through the proximity theorem.
% Similar results
%hold for the zero point energy of a  massless
%scalar field satisfying Dirichlet or Neumann boundary conditions.

\end{abstract}

%\pacs{PACS number(s): 42.50.Lc, 42.50.Dv, 12.20.-m}

\maketitle

\section{Introduction}

The existence of an attractive force between two uncharged,
perfectly conducting parallel plates was predicted by Casimir more
than fifty years ago\cite{Casimir}. Such force has recently been
measured at the 15\% precision level using cantilevers
\cite{roberto}. A similar force between a conducting plane and a
sphere  has also been measured with progressively higher precision
in the last years using torsion balances \cite{Lamoreaux}, atomic
force microscopes \cite{Mohideen}, and capacitance bridges
\cite{chan1}. As described in Ref.\cite{chan1,chan2}, Casimir
forces may be relevant in nanotechnology. The increasing
experimental precision revives the interest in the theoretical
calculation of Casimir forces and energies for different
geometries (see Ref.\cite{Bordag} for a recent review of
experimental and theoretical developments). Only for a few
geometries the exact results are known: parallel plates
\cite{Casimir}, self-energy for spheres \cite{boyer,piro,bowers}
and cylinders \cite{deraad,nesterenko,romeo}. For the force
between a plane and a sphere the exact result is not known, and
only an estimation valid when both are close enough is available.
This estimation is based on the so called proximity theorem
\cite{derjaguin}. Similar estimations have been recently obtained
for the force between two spheres \cite{schaden1}. The problems of
two concentric spheres \cite{saha} and two concentric cylinders
\cite{sahats} have also been considered recently. For both cases
expressions for the Casimir energy have been obtained using the
Abel-Plana formalism.  However a detailed numerical calculation
and analysis of the results is still missing.

In this paper we compute the Casimir energy
for two perfectly conducting and  concentric cylinders,
using  approximate semiclassical methods, and compare the results
with those of an exact calculation based on a  mode-by-mode
summation method.
We will show that the semiclassical approximation describes
accurately the
interaction energy between the cylinders far  beyond the range of validity
of the proximity theorem.

Let us consider a system of conducting shells $S_a$, and denote by
$\Lambda$ a suitable set of parameters describing the geometry
of the system.
For convenience the system is enclosed into a large box, whose boundary
$\Sigma$ will eventually be removed to infinity.
The Casimir energy can be formally defined as
\cite{aclaracion}
\begin{equation}
E_c=E_0(\Lambda)-E_0(\Lambda_0)\,\, ,
\label{def}
\end{equation}
where $E_0 (\Lambda)$ denotes the zero point energy of the electromagnetic
field for the geometry under consideration, and
 $E_0 (\Lambda_0)$ is the one corresponding to the reference
vacuum. For semiclassical calculations it will be enough
to take
the
interior of $\Sigma$ without additional conductors as the
reference vacuum \cite{bd2}. However, for the exact calculation
we will use Eq.(\ref{def}) with $\Lambda_0$ describing
a system in which the shells are very far away.

In terms of the modes of the electromagnetic
field, the  Casimir energy  can be written as
\begin{equation}
E_c={\hbar \over 2}\sum_p(w_p-\tilde w_p)\,\,\, ,
\label{ecasmodes}
\end{equation}
where $w_p$ are the eigenfrequencies of the electromagnetic
field satisfying perfect conductor boundary conditions on $S_a$,
and $\tilde w_p$ are those corresponding
to the reference vacuum.
The subindex $p$ denotes the set of quantum numbers associated to each
eigenfrequency.

For some symmetric systems the Casimir energy can be computed
from the sum over modes, Eq.(\ref{ecasmodes}).
The mode-by-mode summation technique introduced in Ref.\cite{piro} is
based on the use of Cauchy's
theorem to convert the sum over eigenfrequencies into a
contour integral, and it turns out to be a very efficient method of calculation in this context.

\begin{figure}[ht]
\begin{center}
\includegraphics[width=16cm,height=16cm]{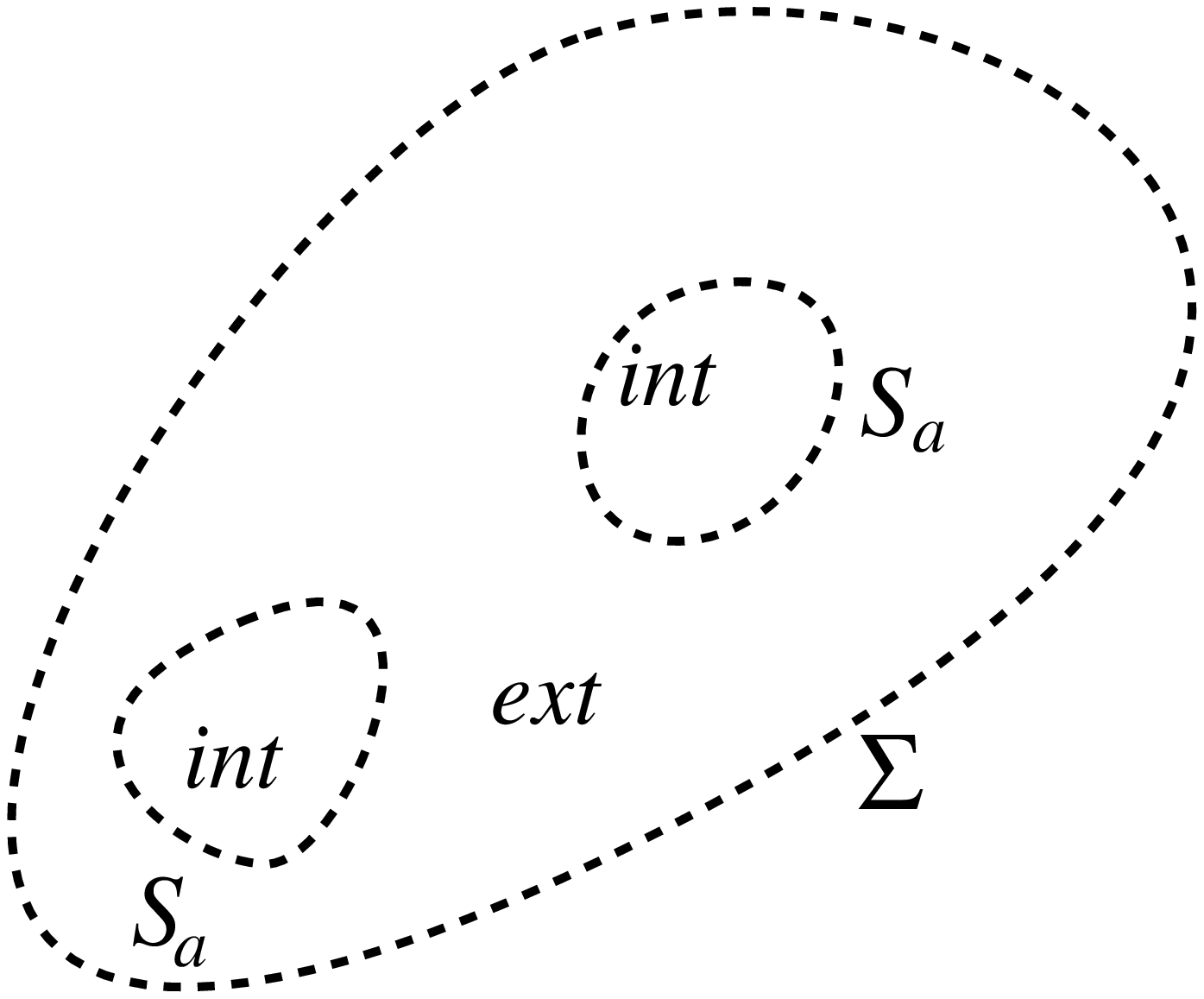}
\caption{Conducting shells $S_{a}$, not necessarily connected, limiting
internal and external regions,  with the
space cut-off $\Sigma$ limiting the external region.} \label{fig1}
\end{center}
\end{figure}

Alternatively, the Casimir energy can be computed from
the knowledge of the density of electromagnetic modes
$\rho (E)$ inside $\Sigma$. Of course this is as difficult
as computing the sum over modes. However, semiclassical
estimates for $\rho (E)$ allow for semiclassical
approximations for the Casimir energy \cite{schaden1,schaden2}.
The Casimir energy
can be written as
\begin{equation}\label{ec}
 E_{c} = \int_{0}^{\infty} \frac{1}{2} \;  E \;  \Delta \rho (E) \; dE \; ,
\end{equation}
where
\begin{equation}\label{ros}
  \Delta \rho (E) \equiv \sum_a \rho_{int,a} + \rho_{ext} - \rho_{vac}  \; ,
\end{equation}
describes the change in  the density of electromagnetic modes inside the box
$\Sigma$ when the conducting shells $S_a$ are introduced.
The spectral density in each region $r = int, ext$ or  $vac$
(see Fig.\ref{fig1}),
$\rho_{r}(E) = \sum_n \delta(E - \hbar \omega_{r_{n}})$, is given in terms of
a sum on the respective  eigenfrequencies $\omega_{r_{n}}$.

Given the Casimir energy, one can compute the forces acting on the
shells $S_a$ by taking appropriate derivatives with respect to
their positions. For the particular case we will consider in this
paper, namely two infinite concentric cylinders of radii $a$ and
$b$ with $a<b$, it is easy to see using symmetry considerations
that the net force on each shell vanishes. However, the pressure
is different from zero. For example, the pressure on the inner
shell is given by
\begin{equation}
p_c=-\frac{1}{2\pi a\ell}\frac{\partial E_c}{\partial a}
\label{pres}
\end{equation}
where $E_c$ is the Casimir energy for  a length $\ell$.

In Ref.\cite{schaden1,schaden2}, the Casimir energy for certain systems has been evaluated
semiclassically using the fact that in the limit $\hbar\rightarrow 0$,
$\Delta \rho (E)$
can be computed as a sum over periodic orbits of a free particle
bouncing off the surfaces $S_a$.
This approximation has been
applied to (chaotic) non-symmetric
configurations and the contribution of  only one isolated
periodic trajectory has been considered.
This is the case for the force between a plane and a sphere, or for the force
between  non-concentric spheres. Here we will extend the formalism to
symmetric configurations (integrable cavities) in which a whole family of
non isolated periodic trajectories contribute
significantly in the semiclassical limit.

The paper is organized as follows. In Section \ref{s1}, for those
readers not acquainted with semiclassical methods, we present an
overview of periodic orbit theory and semiclassical
approximations.  Section \ref{s2} is  devoted to  the
semiclassical calculation of Casimir interaction between two
coaxial cylindrical conductors. In Section \ref{sex} we compute
the exact Casimir energy for the same system, using the mode by
mode summation technique. In Section \ref{snum} we present a
detailed comparison between the exact, the semiclassical, and the
proximity theorem  results. Section \ref{scon} contains the main
conclusions of our work.

\section{Overview of semiclassical methods and  Periodic orbit theory:
General Formalism}\label{s1}

Periodic orbit theory relates oscillations in the quantum level density of a
given Hamiltonian to the periodic orbits in the corresponding
classical system.
It has its origin in the well known decomposition of the spectral density
as
\begin{equation}\label{rosemi}
\rho(E)= \hat \rho(E) \, + \rho^{osc} (E);
\end{equation}
that has a rigorous meaning only in the semiclassical
$(\hbar \rightarrow 0)$
regime for which the scales of variation of $\hat \rho(E)$ and
$ \rho^{osc} (E)$ decouple.

To leading order in $\hbar$ the smooth term $\hat \rho(E)$ is the
Thomas-Fermi contribution  that takes into account the volume of
accessible classical phase space at energy $E$. For non
relativistic particles confined in cavities with reflecting walls,
corrections to $\hat \rho(E)$ of higher order in $\hbar$ were
first given by Weyl \cite{w}. Later Balian and Bloch \cite{bb1}
derived the  generalized Weyl formula for cavities with arbitrary
smooth convex boundaries  and for general classes of boundary
conditions. We refer the reader to the book of Baltes and Hilf
\cite{bal} for many details and examples on the Weyl's expansion.

The deviations from the smooth behavior are given by the  oscillating contribution
 $ \rho^{osc} (E)$, which might be written as a sum over classical periodic orbits.
The foundations of periodic orbit theory (POT) are closely related to the
semiclassical quantization of integrable systems introduced by Bohr and
Sommerfeld.
Using a multiple reflection expansion for the time independent Green function
and employing the principle of stationary phase, Balian and Bloch \cite{bb2}
derived in 1972 a trace formula for cavities with ideal reflecting walls giving
 explicit results for spherical cavities. Later Berry and Tabor
\cite{bt} obtained exactly  the same  formula for integrable cavities using
the torus
quantization rule and  the Poisson summation formula to express the
spectral density
$\rho (E)$ as a multiple integral  and evaluating the integrals by the
saddle point method.

The  Bohr-Sommerfeld quantization rule fails for
quantum Hamiltonian systems whose classical counterpart displays
chaotic motion.  In 1971 Guztwiller presented his famous trace
formula  (see Ref. \cite{gutz} for a review) obtained from
Feynman's path integrals and based on the semiclassical expansion
for the single particle propagator. The trace formula obtained by
Gutzwiller is only applicable if all the involved periodic orbits
are isolated in phase space, being for this reason particularly
successful in its applications to chaotic systems. Nevertheless it
fails for systems which have degenerate families of
 non-isolated periodic orbits, such as integrable systems.

POT has been extensively used during the last twenty years to obtain
semiclassical estimates for  energy spectra of quantum (many-body) systems
like atoms and  nuclei, with vast  applications in  nuclear and atomic physics.
More recently in condensed matter physics, several transport properties
 of mesoscopic systems like metal clusters, metallic grains, quantum dots and
microstructures have been studied employing  the semiclassical tools  \cite{brack}.

In  recent works the semiclassical Gutzwiller trace
formula was exploited to evaluate the leading order Casimir energy  for
some  ``chaotic'' configurations of  pair of conductors, like two conducting
spheres separated a small distance apart
or an open shell and a
sphere \cite{schaden1,schaden2}.
Although the authors of these works comment on the impossibility to employ
Gutzwiller method to evaluate the Casimir energy when non-isolated periodic
orbits are present, they did not discuss the possible use of the Balian-Bloch (BB) or
Berry-Tabor (BT)
trace formulae. As we will show below,  they can be properly adapted  to
evaluate the Casimir effect for   integrable configurations of conductors
({\it i.e.} spherical shells, cylinders, parallel plates, etc.)

Our approach will be to employ  a modified version of
the BT trace formula \cite{bt}, that provides  the
appropriate path to evaluate semiclassically the  Casimir effect for
the case of  two coaxial cylindrical conductors.

We should also mention that Balian and Duplantier\cite{bd} obtained the distribution of electromagnetic modes inside a conducting cavity  using a
multiple scattering  expansion for the  electromagnetic
Green functions, that leads to a decomposition of the spectral density
in a smooth and an  oscillatory term  that can be  written as a sum
over closed polygons. In a subsequent work they  tested the convergence
of their method evaluating  the Casimir
energy for a perfectly conducting sphere \cite{bd2}.

$\rho^{osc}$ will turn to be  the main
ingredient in the semiclassical evaluation of the Casimir energy
Eq. (\ref{ec}) and can be  written,
to leading order in $\hbar$, as a sum \cite{gutz}
\begin{equation}
\rho^{osc} (E)= \frac{1}{\hbar^\nu} \sum_t \; A_{t}(E) \;  \sin ( S_{t}(E)/ \hbar
 \; + \mu_{t} ) \; ,
\end{equation}
running over periodic orbits  labeled by $t$, where $S_{t}$ is the classical
 action of the periodic orbit  $t$ whose period is $T = d S_{t}/ d E $.
For  photons of energy $E$, the dispersion relation
is $E = p \; c $,  being  $p$ the momentum and
$c$ the velocity of light.  The classical action along a periodic
orbit $t$ is then
\begin{equation}
S_{t}(E)= \oint {\bf p} . \; d{\bf q} = p \; L_{t} = \frac {E}{c} \; L_{t} \; ,
\end{equation}
with $L_{t}$ the length of the periodic orbit.

The phase $\mu_{t}$ is the so-called Maslov index that counts
the number of caustics along the periodic orbit. One should
distinguish between Dirichlet and Neumann boundary conditions. For
Dirichlet boundary conditions, an additional phase $\pi$ should be
taken into account in $\mu_{t}$ for each bounce of the trajectory
on the walls of the cavity.

The exponent $\nu$ and the amplitudes  $A_{t}$ depend on the type of
periodic orbit.  For integrable systems with $d$  degrees of
freedom, the periodic orbits
are non-isolated
and form $(d -1)$-parameter families, with the  resulting exponent $\nu = (d \; + 1 )/2 $. In chaotic
systems the orbits are isolated and their contribution is semiclassicaly
smaller with $\nu =1$, irrespective of the dimensionality.

\section{Semiclassical results for
two coaxial cylindrical conductors} \label{s2}

We  consider two infinite and concentric cylindrical conducting shells
$C_1$ and $C_2$, of radii $a$ and $b$ (with $a<b$).
Let us call    $\rho_{12}$ the spectral  density for
photons confined in the region  between
$C_1$ and $C_2$ and $\rho^{osc}_{12}$ the
corresponding leading order oscillating contribution.
To obtain  $\rho^{osc}_{12}$  one can derive  {\it \` {a} la} BT a
 trace formula for photons
inside the  conducting shells,  starting from  the  Hamiltonian
$H = E= p \; c$, employing Poisson summation formula
and using  the torus quantization rule \cite{majo}.
However we find it easier to  proceed in two steps.
Firstly,  if
one knows the oscillating contribution $\rho^{osc} (E_m)$
for non-relativistic particles of mass $m$ and energy $E_m = {p_m} ^2 /2m$
confined in a given cavity,
the oscillating contribution for photons confined in the same  cavity
is obtained from  $\rho^{osc} (E_m)$ after replacing
$E_{m} \rightarrow p/c $ and $ p_{m}/ m \rightarrow c$.
This is not restricted to the oscillating contribution, and the same
replacement could be done in the complete spectral density $\rho$.
Secondly, for cavities with axial symmetry, and in particular for the
geometry under consideration, the periodic orbits (PO) are
contained in planes perpendicular to the $z$ axis (we are considering
cylinders of  infinite length). Therefore if one knows the oscillating
contribution to the spectral density in the bidimensional annular region
between two  disks, $\rho^{osc}_{\odot}$, it its straightforward to
obtain\cite{majo} that, for a given length $\ell$,
\begin{equation}\label{2d3d}
\rho^{osc}_{12} (E = \hbar c \; k ) = \frac{\ell}{\hbar \; c \pi} \;
\int_{0}^{k} \frac {k}{\sqrt{k^2 - k_{z}^2}} \; \rho^{osc}_{\odot}(\sqrt{k^2 - k_{z}^2}) \; d k_z \;.
\end{equation}
We stress  that the  above equation is valid
for cavities with  axial symmetry, but with generic  transverse sections.

\begin{figure}[ht]
\begin{center}
\includegraphics[width=11.5cm,height=12cm]{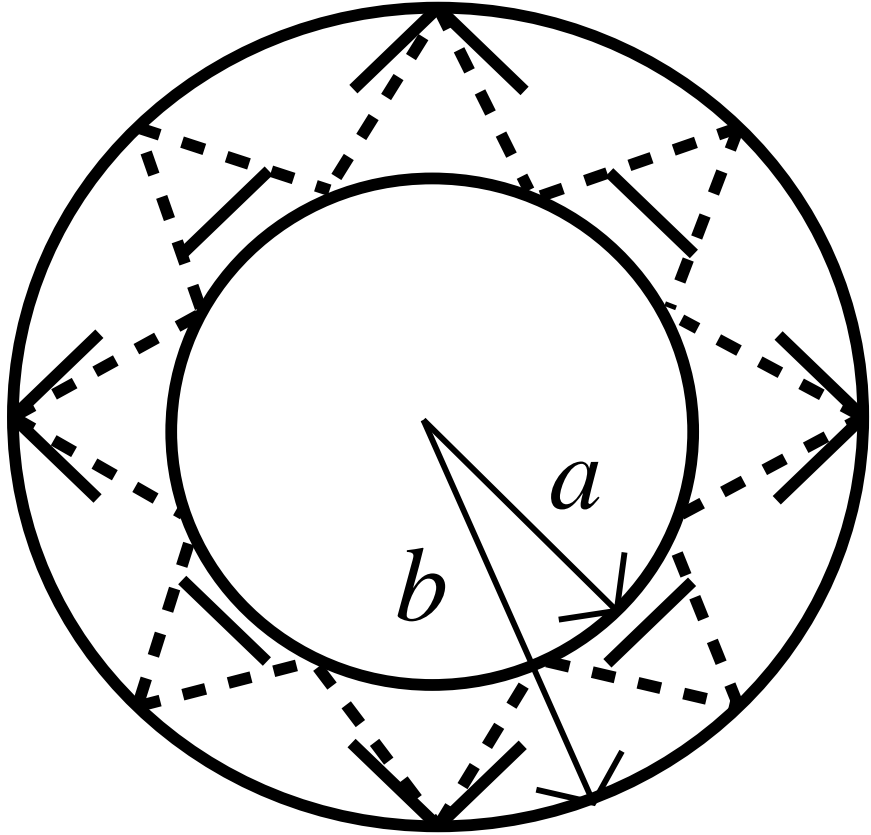}
\caption{Two periodic orbits in a bidimensional annular region.
The type-I orbit $(v=4, w=1)$  does not touch the inner disk
(long-dashed line). The type-II orbit  $(v=8, w=1)$  hit it (short-dashed line). } \label{fig2}
\end{center}
\end{figure}

The oscillating contribution   for  non-relativistic particles confined in
bidimensional annular regions, $\rho^{osc}_{\odot} (E_m)$
has been previously derived (see e.g. Richter's book in Ref.\cite{brack}
and Appendix A).
In this configuration there are two types of PO (see Fig. \ref{fig2}):
those
which do not touch the inner
disk   (type-I), and those which do hit it (type-II).

The type-I orbits are polygons that may be uniquely labeled by two integers
$(v, w)$ where $v$ is the number of vertices (or sides) and $w$ the winding
number around the center. For $w =1$ one has ordinary regular polygons, and
for $w >1$ star-like polygons, except if $v$ and $w$ are coprimes.
In this situation the label $(v, w) =( n \; j, n \;  k)$ describes the $n^{th}$
repetition of the primitive orbit with $j$ vertices and winding number $k$.
Type-I orbits must fulfill
$v \ge \hat v (w) \equiv Int [w \pi / \arccos (a/b)]$, that is the minimum
value of $v$ depends on  $w$ and the ratio  $ a/b$ between the
radii of the inner and outer cylinders, respectively. This is  a geometrical
 restriction due to the fact that $\cos ( \pi w/v) > a/b$.
The lengths $L_{vw}$ of these orbits are
\begin{equation}\label{lI}
L_{vw} = 2 \, v \; b \sin (\pi w /v) \; .
\end{equation}
Type-II trajectories are labeled by $(v, w)$, where now $w$ is the number
of turns around the inner disk in order to come back to the initial point after
$v$  bounces on the outer circle of radius $b$. We have the same restriction
$v \ge \hat v (w)$ as for type-I trajectories
and the length is given by
\begin{equation}\label{lII}
\bar L_{vw} = 2 \, v \; b \;  \sqrt{ \left( 1+ \; \left(\frac{a}{b}\right)^2
\; - 2 \; \frac{a}{b}  \; \cos (\pi w /v) \right)} \; .
\end{equation}

Following the steps described above, we write $\rho^{osc}_{12} \equiv \rho^{osc}_{12,I} +
\rho^{osc}_{12,II}$ as
a  sum of two terms that take into account the contributions  of
type-I and type-II  PO, respectively. We find (see Appendix A)
\begin{eqnarray}
\rho^{osc}_{12,I} (E) & = &  \sum_{w \ge 1} \;  \sum_{v \ge \hat v}
\frac{\ell}{\pi (\hbar c)^2} \; \frac{L_{vw}}{v^2} \; E \; \sin \left(\frac{E}
{\hbar c} \;  L_{vw} \pm \frac{ v \;  \pi}{2} + \frac{\pi}{2} \right) \; ,
\label{roI} \\
\rho^{osc}_{12,II} (E) & = &   \sum_{w \ge 0} \; \sum_{v \ge \hat v}  f_{vw}
\; \frac{ 2 \ell}{\pi (\hbar c)^2} \frac{b^2}{\bar L_{vw}} \; A_{vw} \; E \;
\sin \left( \frac{E}{\hbar c} \; \bar L_{vw} \right)  \label{roII} \; .
\end{eqnarray}

The $+ (-)$ in Eq.(\ref{roI}) corresponds to Dirichlet (Neumann) boundary
conditions (bc).
On the other hand, there is no additional  phase and therefore
no difference  between Dirichlet
and Neumann bc in the  type-II contribution, Eq.(\ref{roII}).
The coefficient  $f_{vw}$ is given by $f_{vw} = 2$, except   for the
self-retracing type-II orbit, that has
$w=0$ and $v =2$, for which $f_{2 0 } = 1 $.
In Eq.(\ref{roII}) we have defined
\begin{equation}\label{avw}
A_{vw} \equiv  \sqrt{ \left( 1 - \;
\frac{a}{b} \; \cos (\pi w /v) \right) \; \left( \frac{a}{b} \; \cos (\pi w /v)
 - \left( \frac{a}{b} \right)^2  \right)} \; .
\end{equation}

Having obtained the explicit expression
for the oscillating contribution to the spectral density in the region
between two coaxial cylinders we can use
this expression to estimate the semiclassical contribution
to the Casimir
interaction in that region.  For this purpose, we need to
rewrite Eq.(\ref{ros}) for the present  configuration of conductors.
The internal region is now formed by the two conducting shells.
We take the outer cylindrical shell $C_2$ as the  shell
that limits the internal and external regions. Therefore we write,

\begin{equation}\label{roint}
\rho_{int}= \rho_{1} + \; \rho_{12} \; ,
 \end{equation}
where $\rho_{1}$ is the density of electromagnetic modes inside the
cylindrical shell $C_{1}$.
 $\rho_{12}$ has been already defined and its oscillating contribution
computed in Eqs.(\ref{roI}) and (\ref{roII}).

In order to employ  the semiclassical decomposition of  the
spectral densities in each region $r$, Eq.(\ref{rosemi}), we
derive the smooth contributions $\hat \rho_{r} (E)$.
Starting  from the Weyl expansion for the non-relativistic case
\cite{w,bb2} and following again the steps described above, we
obtain
\begin{equation}\label{sm}
\hat \rho_{r} (E) = \frac{\pi \; E^2}{2 \; (\hbar \; c \; \pi)^3} \; {\cal V}_{r}  \; \mp
\frac{ \pi \; E}{8 \; (\hbar \; c \; \pi)^2} \; {\cal S}_{r} \; + {\cal O}
( 1/ \hbar ) \;,
\end{equation}
where ${\cal V}_{r}$ and ${\cal S}_{r}$ are the volume and the surface area of
each region $r$.
The $- (+)$ sign in the surface term corresponds to Dirichlet (Neumann) bc.

Therefore replacing  Eqs.(\ref{rosemi}), (\ref{roint}) and (\ref{sm}) in
Eq.(\ref{ros}) we obtain,  {\it to leading order in $\hbar$},

\begin{equation} \label{deltarosemi}
\Delta \rho^{sem} (E) = \Delta \hat \rho  (E) \; +  \Delta  \rho^{osc} (E)
= \rho^{osc}_{1} (E)  + \; \rho^{osc}_{12} (E) + \; \rho^{osc}_{ext} (E) -
  \; \rho^{osc}_{vac} (E) \;,
\end{equation}
where the label $sem$ is employed to emphasize the explicit use of
the semiclassical decomposition, Eq.(\ref{rosemi}).
The  cancellation, to leading order in $\hbar$,  of  the net smooth
contribution $\Delta \hat \rho (E)$ is a consequence of the identity
 ${\cal V}_1 \; +  {\cal V}_{12} \; + {\cal V}_{ext} - {\cal V}_{vac} = 0 $

 For the geometry under study,  and when  considering
the total  electromagnetic (e.m.) contribution, it is
satisfied that   $\Delta \hat \rho (E)= 0$ to next-to-leading order in $\hbar$.
This is due to the opposite sign of the
surface term in Eq.(\ref{sm}) for the Dirichlet and Neumann bc.

As already mentioned, in order to compute the Casimir energy
Eq.(\ref{ec}) one has to remove to infinity the boundary $\Sigma$
that enclose the reference vacuum. In this limit
$\rho^{osc}_{ext} (E) \rightarrow 0$
 and $\rho^{osc}_{vac} (E) \rightarrow 0$  since  no PO
survive in the limit $\Sigma \rightarrow \infty$.
Taking this into account
and in  order to compare the semiclassical results with the exact
calculations, we define in analogy to Eq.(\ref{ec})

\begin{equation}\label{esoc}
E^{sem} \equiv \frac{1}{2} \int_{0}^{\infty} E\;   \Delta \rho^{sem} (E) \; dE
 \; =  \frac{1}{2} \int_{0}^{\infty} E\ \left( \rho^{osc}_{1} (E)  + \;
\rho^{osc}_{12} (E) \right) \; dE  \equiv  E^{osc}_{1}
\; + E^{osc}_{12} \; .
\end{equation}

It is a well known result that in the case of cylindrical geometries
the Casimir energy for the electromagnetic case (e.m.)
can be computed as the sum of   the Dirichlet and Neumann scalar
 contributions.
Thus, one has

\begin{equation}\label{esemem}
E^{sem} =\; E^{osc}_{1} \; + E^{osc}_{12} \;=
 E^{osc}_{1,D} \; +  E^{osc}_{1,N} + \; E^{osc}_{12,D}  \;
+ E^{osc}_{12,N}  \; ,
\end{equation}
where the subscript $D (N) $ stands  for Dirichlet (Neumann) bc.

We now turn our attention  to  $\rho^{osc}_{1} (E)$, the oscillating
contribution for the  density of modes inside $C_1$.
We derive $\rho^{osc}_{1} (E)$ adapting the previous
result Eq.(\ref{roI}).
The PO are the same  type-I polygons labeled by two integers
$(v, w)$. Being $a$ the radius of $C_1$, the length of these orbits is now
 ${\tilde L}_{vw} = 2 \, v \; a \sin (\pi w /v) $, the shortest one
corresponds to the self-retracing diametrical orbit that has $v=2, w=1$.
For  a cylindrical shell like $C_{1}$, it is satisfied that
$v \ge 2 w $, that is $\hat v = 2\; w$.
In other words, the cylindrical cavity is the annular one
with a vanishing inner radius. Therefore,
\begin{equation}\label{ro1}
\rho^{osc}_{1,bc} (E)  = \;  \sum_{w \ge 1} \;  \sum_{v \ge 2 w } \; g_{vw}
\; \frac{\ell}{2 \pi (\hbar c)^2} \; \frac{{\tilde L}_{vw}}{v^2} \; E \; \sin \left(\frac{E}
{\hbar c} \;  {\tilde L}_{vw} \pm \frac{ v \;  \pi}{2} + \frac{\pi}{2} \right)
 \; ,
\end{equation}
where again the $+ (-)$ sign corresponds to $bc = D (bc = N)$.
The prefactor is  $g_{21} = 1$ for the self-retracing orbit  and $g_{vw} = 2 $
otherwise.

Taking into account that
\begin{equation}\label{eosc1}
 E^{osc}_{1} = E^{osc}_{1,D} + \;  E^{osc}_{1,N} = \;  \frac{1}{2} \int_{0}^{\infty} E  \left(
\rho^{osc}_{1,D} (E) \; + \rho^{osc}_{1,N} (E) \right) dE \; ,
\end{equation}
 we  perform the energy integration  employing Eq.(\ref{ro1}) and
an exponential cutoff.
It is straightforward to show that
\begin{eqnarray}\label{int}
& & \int_{0}^{\infty} E^2   \sin \left(\frac{E}
{\hbar c} \;  {\tilde L}_{vw} \pm \frac{ v \;  \pi}{2} + \frac{\pi}{2} \right)
\; dE  =  \lim_{ \lambda \rightarrow 0} \int_{0}^{\infty} \exp{ (-  \lambda
\;  E)} \;  E^2   \sin \left(\frac{E}
{\hbar c} \;  {\tilde L}_{vw} \pm \frac{ v \;  \pi}{2} + \frac{\pi}{2} \right)
\; dE   \nonumber \\
& = & \pm 2 \; \left( \frac{ \hbar \; c}{{\tilde L}_{vw}} \right)^3 \; \sin ( \pi \; v /2 ) \; ,
\end{eqnarray}
and in conclusion
\begin{equation} \label{eos1}
E^{osc}_{1} = E^{osc}_{1,D} \; +   E^{osc}_{1,N} = \; 0 \; .
\label{semi1cil}
\end{equation}

Thus in the semiclassical approximation, the e.m. Casimir energy
for an isolated cylinder vanishes. This is also
the case to lowest order in the multiple
scattering expansion \cite{bd2}.

To compute $E^{osc}_{12} = E^{osc}_{12,D} + \;  E^{osc}_{12,N}$
 we should take into  account, in principle, both  type-I and type-II PO for
each scalar contribution, $D$ or $N$. However, as before,  the  contributions
from type-I PO cancel out and only type-II PO
contribute to the semiclassical e.m. Casimir interaction.
We  compute $E^{osc}_{12}$ noting that
$\rho^{osc}_{12,II}$ is the same for Dirichlet and Neumann bc and therefore
\begin{equation}\label{eosc12II}
 E^{osc}_{12} = 2 \;  \frac{1}{2} \int_{0}^{\infty} E  \;
\rho^{osc}_{12,II} (E)  \; dE \;.
\end{equation}

Replacing $\rho^{osc}_{12,II}$ from Eq.(\ref{roII}) and
performing again the energy integration using an exponential cutoff,
we obtain the main result of this section. Namely,

\begin{equation}\label{main}
E^{sem} = \; E^{osc}_{12} = \;
 -  \ell  \frac{\hbar \; c}{ 4 \; \pi \; a^2} \; \alpha ^{1/2}
\; \sum_{w \ge 0} \;  \sum_{v \ge \hat v} f_{vw}  \;
\frac{1} {v^4} \;  N(\alpha, v,w)\,\, ,
\end{equation}
where we have defined $\alpha \equiv b/a$,  and
\begin{equation}\label{nvw}
N(\alpha,v,w) \; \equiv \frac{\sqrt{ \left( \alpha - \;  \cos (\pi w /v) \right) \;
\left( \alpha \; \cos (\pi w /v) - \; 1 \right)}}
{ \left( 1 \; + \alpha ^2 \; - 2 \; \alpha \;  \cos (\pi w /v) \right) ^2} \; .
\end{equation}

It is interesting to rewrite Eq.(\ref{main}) by separating the terms with  $ w = 0$ from
those with $ w  \ge 1 $.
The orbits with  $w =0 $ and $v \ge  1$ are  the diametrical  orbit  and
its repetitions, whose lengths are  $\bar L_{v 0} = 2 \; v \; 
(b\; - \; a) $.
Therefore we write

\begin{equation}\label{maindiv}
E^{sem} \equiv  E^{sem}_{w = 0} \; + E^{sem}_{w \ge 1} =\; E^{sem}_{w=0}  - \;
\ell  \frac{\hbar \; c}{ 2 \; \pi \; a^2} \; \alpha ^{1/2}
\; \sum_{w \ge 1}  \; \sum_{v \ge \hat v} \;
\frac{1} {v^4} \;  N(\alpha, v,w) \; .
\end{equation}

In order to obtain $E^{sem}_{w = 0}$, we  calculate from  Eq.(\ref{nvw})
\begin{equation}\label{nv0}
N(\alpha,v,0) = \frac{1}{( \alpha \; - 1)^3} \; ,
\end{equation}

and using the fact that $\sum_{v \ge 1} \; v^{-4} = \zeta (4) =
{ \pi^4}/ {90} $ we finally obtain,

\begin{equation}\label{ew0}
E^{sem}_{w=0} = \;- \ell \; \frac{\hbar \; c \; \pi^3}{360 \; a^2} \; \frac{\sqrt{\alpha}}
{( \alpha \; - 1)^3} \; = -  \ell  \; \frac{\hbar \; c \; \pi^3}{360} \; \frac{\sqrt{ a \; b}}
{( b \; - a )^3} \; .
\end{equation}

This expression, which is valid for arbitrary values of $\alpha > 1$, diverges as $\alpha \rightarrow 1$. One
might wonder whether the sum of the contributions with $w \ge 1$
cancels and/or contributes, in any way,  to this
divergence. To investigate this problem it is interesting to note that, in
such a limit, it is possible to derive an
analytic expression for the sum of these contributions. Due to the fact that  for  $\alpha \sim 1$,
\begin{equation}
\hat v  \equiv Int [w \pi / \arccos (a/b)] \sim
\frac{ \pi \; w \sqrt{\alpha}}{ \sqrt { 2 \; ( \alpha - \; 1)  }}
 >> \; w \; ,
\end{equation}
we can approximate $\cos (\pi w /v) \sim 1 \; - {(\pi \;  w /v)}^{2} /2 $.
Thus, from Eq.(\ref{nvw})
\begin{equation}\label{nvwal1}
N(\alpha \sim 1 ,v, w) = \;  \frac{1}{( \alpha \; - 1)^3} \;
\frac{ v^4 - \frac{ (\pi \; w )^4}{ 16 \; (\alpha - \; 1 )^2}}
{\left( v^2 + \frac{ (\pi \; w )^2 \; \alpha}{ (\alpha - \; 1 )^2 } \right)
 ^2} \; .
\end{equation}
Since $\hat v >> 1$ we can convert the sum $\sum_{v \ge \hat v} \;
\frac{1} {v^4} \;  N(\alpha \sim 1, v,w)$ in Eq.(\ref{maindiv})
into an integral.
Performing the integration and summing over $w \ge 1$ we  obtain
\begin{equation}
E^{sem}_{w \ge 1} (\alpha \sim 1) = - \;
\ell  \frac{\hbar \; c}{ 4 \; \pi^3  \; a^2} \frac{\zeta (3)}{\alpha}\;
+ {\cal O} (\alpha \; - 1),
\end{equation}
which implies that $E^{sem}_{w \ge 1}$ remains finite as $\alpha \rightarrow 1$. Thus, in this limit
the semiclassical result is divergent and completely dominated by the $w=0$ contribution.
As we will see later, this dominance extends over a
rather large range of values of $\alpha$.

Having shown that for $\alpha \sim 1$ one has $E^{sem} \sim
E^{sem}_{w =0}$ it is interesting to compare Eq.(\ref{ew0}) with
what results from the application of the proximity theorem
\cite{derjaguin}. The Casimir energy  between two parallel plates
of area $A$ separated a small distance $\delta$ is
\begin{equation} \label{paral}
E_{||} = - \;  \frac{\hbar \; c \; \pi^2}{720} \frac{A}{\delta^3} \,\, .
\end{equation}
For the two coaxial cylinders, when
$( b \; - a )\ll a\sim b$,
the proximity approximation for the
Casimir energy, $E_P$,  is obtained  taking $\delta \; = ( b \; - a )$
and  $A \rightarrow 2 \; \pi \ell
\; a \sim 2 \; \pi \; \ell b \;$
in Eq.(\ref{paral}). The result is
\begin{equation}\label{derj}
E_{P} = \left \{ \begin{array} {ll}
- \ell \; \frac{\hbar \; c \; \pi ^3}{360}  \; \frac{a}{(b \; - a)^3} &
\mbox{for $ \;  A = 2 \; \pi \ell \; a$ ,} \\
- \ell \; \frac{\hbar \; c \; \pi ^3}{360} \; \frac{a \; \alpha}
{(\alpha \; - 1)^3} &
\mbox{for $ \;  A = 2 \; \pi \ell \; b. $}
\end{array} \right.
\label{ED}
\end{equation}
The proximity theorem does not suggest any particular
choice for the area A. This is irrelevant in the limit
$\alpha \sim 1$.

The semiclassical result Eq.(\ref{ew0})
can be interpreted as a proximity approximation
with a given choice for the area: the geometric mean
between  the areas of both cylinders.
Although it has been shown that to compute a proximity force for a
 ``chaotic" geometry,  the
geometric mean radius should be taken \cite{blocki},
such demonstration does not apply to the present case. However,
the semiclassical approximation provides a justification for it.
As we will see in
Section V, this choice
reproduces the exact  pressure with an error less than $10 \%$
for $1<\alpha <4$, improving
the proximity result beyond its expected range of validity.

\section{Exact Casimir energy}\label{sex}
In this section we will compute the exact Casimir energy for the
coaxial cylinders using the mode by mode summation method
of Ref.\cite{piro}.

The zero point energy in Eq.(\ref{ecasmodes}) is ill defined,
because both series are
divergent. In order to define it properly we introduce a cutoff function
as follows
\begin{equation}
E_{ex}(\sigma)={\hbar\over 2}\sum_p(e^{-\sigma w_p} w_p-e^{-\sigma \tilde w_p}
\tilde w_p) \; .
\label{ecasreg}
\end{equation}
The exact Casimir energy $E_{ex}$ is the limit of $E_{ex}(\sigma)$ as
$\sigma\rightarrow 0$.
For simplicity we choose an exponential cutoff, although the explicit
form is not relevant. In our definition we take $w_p$ as the eigenfrequencies of the electromagnetic
field satisfying perfect conductor boundary conditions at $r=a, r=b$ and $r=R$,
and $\tilde w_p$ are those corresponding to the boundary conditions
at $r=R_1, r=R_2$ and $r=R$, in the limit $R>R_2>R_1\gg a > b$.
 $R_1, R_2$ and $R$ are the parameters that define the reference vacuum
(see Eq. (\ref{def})). $R$ corresponds in this case to  the radius of the
surface $\Sigma$ in Fig.\ref{fig1}.

In cylindrical coordinates, the eigenfunctions are of the form
\begin{equation}
h_{n k_z}(t,r,\theta,z)=e^{(-iw_{n k_z}t+ik_z z+in\theta)}R_n(\lambda r) \; ,
\label{eigenf}
\end{equation}
where the function $R_n$ is a combination of Bessel functions
satisfying the perfect conductor boundary conditions. These boundary
conditions define the possible values of the constant
$\lambda$. The
eigenfrequencies are $w_{n k_z}=c \sqrt{k_z^2+\lambda^2}$.

The frequencies of the TE modes are defined by the following relations:
\begin{eqnarray}
&& J_n(\lambda a)=0\,\,\,\,\,\,\,\,\,\,\,\,\,\,\,\,\,\,\,\, r<a\nonumber\\
&&J_n(\lambda a)N_n(\lambda b)-J_n(\lambda b)
N_n(\lambda a)=0\,\,\,\,\,\,\,\, a<r<b\nonumber\\
&&J_n(\lambda b)N_n(\lambda R)-J_n(\lambda R)
N_n(\lambda b)=0\,\,\,\,\,\,\,\, b<r<R \; .
\label{condtm}
\end{eqnarray}
For later use we introduce the notation $F_n^{TE}$ for the
product of these three relations
\begin{equation}
F_n^{TE}(z,a,b,R)=J_n(z a)[J_n(z a)N_n(z b)-J_n(z b)
N_n(z a)][J_n(z b)N_n(z R)-J_n(z R)
N_n(z b)] \; .
\end{equation}

The frequencies of the TM modes involve derivatives of the Bessel
functions:
\begin{eqnarray}
&& J'_n(\lambda_{nm}a)=0\,\,\,\,\,\,\,\,\,\,\,\,\,\,\,\,\,\,\,\, r<a\nonumber\\
&&J'_n(\lambda_{nm}a)N'_n(\lambda_{nm}b)-J'_n(\lambda_{nm}b)
N'_n(\lambda_{nm}a)=0\,\,\,\,\,\,\,\, a<r<b\nonumber\\
&&J'_n(\lambda_{nm}b)N'_n(\lambda_{nm}R)-J'_n(\lambda_{nm}R)
N'_n(\lambda_{nm}b)=0\,\,\,\,\,\,\,\, b<r<R \; .
\label{cond2c}
\end{eqnarray}
As before, we introduce the notation
\begin{equation}
F_n^{TM}(z,a,b,R)=J'_n(z a)[J'_n(z a)N'_n(z b)-J'_n(z b)
N'_n(z a)][J'_n(z b)N'_n(z R)-J'_n(z R)
N'_n(z b)] \; .
\end{equation}

The set of quantum numbers $p$ in Eq. (\ref{ecasreg}) is given by
$(n,m,k_z)$, where $m$ denotes the different solutions
$\lambda_{nm}$ of both $F_n^{TE}(z,a,b,R)=0$ and $F_n^{TM}(z,a,b,R)=0$ .
As we are considering
infinite cylinders, we can
replace the sum over $k_z$ by an integral. The result for the
Casimir energy for a  length $\ell$ is
\begin{equation}
E_{ex} (\sigma) = {\ell \hbar c
\over 2}\int_{-\infty}^{\infty}{dk_z\over 2\pi}
\sum_{n,m}\left (\sqrt{k_z^2+\lambda_{nm}^2}e^{-\sigma c
\sqrt{k_z^2+\lambda_{nm}^2}}
- \sqrt{k_z^2+\tilde \lambda_{nm}^2}
e^{-\sigma c \sqrt{k_z^2+\tilde\lambda_{nm}^2}}
\right) \; ,
\label{exs}
\end{equation}
where we have denoted with $\tilde\lambda_{nm}$ the solutions of
the equations  $F_n^{TE}(z,R_1,R_2,R)=0$ and $F_n^{TM}(z,R_1,R_2,R)=0$.

Using Cauchy's theorem it follows that
\begin{equation}
{1\over 2\pi i}
\int_C \,dz \;  z \; e^{-\sigma z} {d\over dz}\ln f(z)=\sum_i x_i \;
e^{-\sigma x_i} \; ,
\end{equation}
where $f(z)$ is an analytic function within the
closed contour $C$, with simple zeros at $x_1,x_2,...,$ within $C$.
Following Refs. \cite{piro,bowers}, we use this result to replace the sum
over $m$ in Eq.(\ref{exs}) by a contour integral
\begin{equation}
E_c (\sigma)={\ell\hbar c\over 4\pi i}\int_{-\infty}^{\infty} {dk_z\over 2\pi}
\sum_n\int_C dz \sqrt{k_z^2+z^2}
e^{-\sigma c \sqrt{k_z^2+z^2}}{d\over dz}\ln F^{EM}_{n\ 2cyl}(z,a,b) \; ,
\end{equation}
where
\begin{equation}
F^{EM}_{n\ 2cyl}(z,a,b)= \lim_{R_{1},R_{2},R \rightarrow\infty}
{F_n^{TE}(z,a,b,R) F_n^{TM}(z,a,b,R)\over F_n^{TE}(z,R_1,R_2,R)
F_n^{TM}(z,R_1,R_2,R)} \; .
\end{equation}

It proves to be convenient to compute the difference between the energy of
the system of two concentric cylinders and the energy of two isolated
cylinders of radii $a$ and $b$
\begin{equation}
E_{12}(\sigma)=E_c(\sigma) - E_1(\sigma,a)-E_1(
\sigma,b) \; .
\end{equation}
The energy of isolated cylinders has been computed previously
using other regularizations
\cite{deraad,nesterenko}. The formal expression
with the cutoff can be
obtained adapting the two cylinders case above. It is given
by
\begin{equation}
E_1(\sigma,a) ={\ell\hbar c\over 4\pi i}
\int_{-\infty}^{\infty} {dk_z\over 2\pi}
\sum_n\int_C dz \sqrt{k_z^2+z^2}e^{-\sigma c\sqrt{k_z^2+z^2}}{d\over dz}
\ln F^{EM}_{n \ 1cyl}(z,a) \; ,
\end{equation}
where
\begin{eqnarray}
F^{EM}_{n \ 1cyl}(z,a)= \lim_{R_{1},R \rightarrow\infty}
{F_n^{(1)TE}(z,a,R) F_n^{(1)TM}(z,a,R)
\over F_n^{(1)TE}(z,R_1,R) F_n^{(1)TM}(z,R_1,R)}
\end{eqnarray}
and
\begin{eqnarray}
F_n^{(1)TE}(z,a,R)&=&J_n(z a)[J_n(z a)N_n(z R)-J_n(z R)N_n(z a)]
\nonumber\\
F_n^{(1)TM}(z,a,R)&=&J'_n(z a)[J'_n(z a)N'_n(z R)-J'_n(z R)N'_n(z a)] \; .
\end{eqnarray}

Therefore, the explicit form for $E_{12}$ is
\begin{equation}
E_{12}(\sigma) = {\ell\hbar c\over 4\pi i}\int_{-\infty}^{\infty}
{dk_z\over 2\pi} \sum_n\int_C dz \sqrt{k_z^2+z^2}
e^{-\sigma c \sqrt{k_z^2+z^2}}{d\over dz}\ln F_{n \ 12}(z,a,b) \; ,
\end{equation}
where
\begin{equation}
F_{n \ 12}(z,a,b)= { F^{EM}_{n \ 2cyl}(z,a,b) \over F^{EM}_{n \ 1cyl}(z,a)
F^{EM}_{n \ 1cyl}(z,b)} \; .
\end{equation}

\begin{figure}[ht]
\begin{center}
\includegraphics[width=10cm,height=10cm]{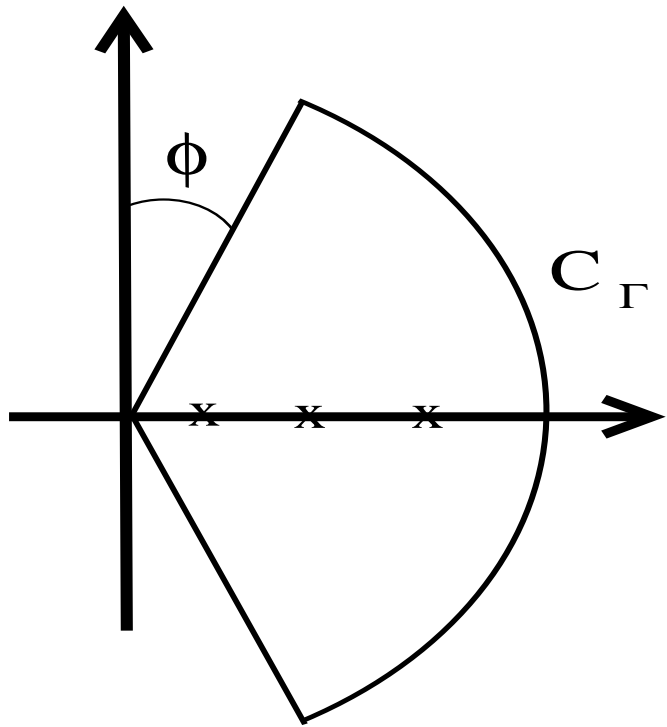}
\caption{Contour for the integration in the complex plane.} \label{fig3}
\end{center}
\end{figure}

To proceed we must choose a contour for the integration in the
complex plane. In order to compute $E_{ex} (\sigma), E_1(\sigma,a)$
and $E_1(\sigma,b)$ separately, an adequate contour is
\cite{bowers} a circular
segment $C_{\Gamma}$ and two straight line segments forming an
angle $\phi$ and $\pi -\phi$ with respect to the imaginary axis (see
Fig. \ref{fig3}).
The nonzero angle $\phi$ is needed to show that
the contribution of $C_{\Gamma}$ vanishes in the limit
$\Gamma\rightarrow \infty$ when $\sigma > 0$. For the rest
of the contour, it can be shown
that the divergences
in $E_{ex}(\sigma)$ are cancelled out by those of $E_1(\sigma,a)$
and $E_1(\sigma,b)$. Therefore
in order
to compute $E_{12}(\sigma)$ we can set
$\phi=0$ and $\sigma =0$.
The contour integral reduces to an integral
on the imaginary axis. We find
\begin{equation}
E_{12}= -{\ell\hbar c \over 2\pi a^2}\int_{-\infty}^{\infty} {dk_z\over 2\pi}
\sum_n Im \left\{\int_0^{\infty} dy \sqrt{k_z^2-y^2}
{d\over dy}\ln F_{n \ 12}(iy,1,\alpha)\right\} \; ,
\label{xx}
\end{equation}
where we have used that $F_{n \ 12}(iy/a,a,b)= F_{n \ 12}(iy,1,\alpha)$.

Using the asymptotic expansions of  Bessel functions
we obtain, on the imaginary axis \cite{aclaracion2}
\begin{eqnarray}
F^{EM}_{n \ 2cyl}(iy,1,\alpha)&=& 16\alpha^2 y^4 I_n(y)
I'_n(y)[I_n(y)K_n(\alpha y)-I_n(\alpha y)K_n(y)]  \nonumber \\
                          &                 &[I'_n(y)K'_n(\alpha y)-
I'_n(\alpha y)K'_n(y)]K_n(\alpha y)K'_n(\alpha y) \\
\label{algo2}
F^{EM}_{n \ 1cyl}(iy,1)&=&-4 y^2 I_n(y)I'_n(y)K_n(y)K'_n(y)   \\
\label{algo}
F_{n \ 12}(iy,1,\alpha)&=&\left[1-{I_n(y)K_n(\alpha y)\over
I_n(\alpha y)K_n(y)}\right] \left[1-{I'_n(y)K'_n(\alpha y)
\over I'_n(\alpha y)K'_n(y)}\right]\,\, .
\end{eqnarray}

Note that $F_{n \ 12}(iy,1,\alpha)$ is a real function. Therefore,
the integral over $y$ in Eq. (\ref{xx}) is restricted
to $y > k_z$. We rewrite Eq.(\ref {xx}) as
\begin{equation}
E_{12}=-{\ell\hbar c\over 2 \pi^2 a^2}\int_0^{\infty}
dk_z\sum_n\int_{k_z}^{\infty}dy \sqrt{k_z^2-y^2}
{d\over dy}\ln F_{n \ 12}(iy,1,\alpha)\; .
\end{equation}
After some trivial steps we obtain
\begin{eqnarray}
E_{12}&=& {\ell\hbar c\over 4\pi a^2}
\sum_n\int_{0}^{\infty} dy \ y\ln F_{n \ 12}(iy,1,\alpha) \nonumber \\
&=&{\ell\hbar c\over 4\pi a^2}\left(\int_{0}^{\infty} dy
\ y\ln F_{0 \ 12}(iy,1,\alpha)+
2\sum_{n=1}^{\infty}\int_{0}^{\infty} dy \ y\ln F_{n \ 12}(iy,1,\alpha)
\right) \; .
\label{xxx}
\end{eqnarray}
The exact energy for the two concentric
cylinders is the sum of $E_{12}$ and the Casimir energies for
single cylinders of radii $a$ and $b$ \cite{nesterenko,romeo}. The final
result is

\begin{equation}
E_{ex} = E_{12} - 0.01356 \; (\frac{1}{a^2}+\frac{1}{b^2}) \;  \ell\; \hbar \;
 c \; .
\label{exfin}
\end{equation}

In the limit $\alpha\sim 1$, $E_{12}$ dominates the Casimir energy
and can be evaluated analytically. Using the uniform expansion
for the Bessel functions we obtain, to leading order in $(\alpha -1)$
\begin{equation}
 F_{n \ 12}(iy,1,\alpha)
\approx \left[1-{\rm e}^
{-2nh(z)(\alpha -1)}\right]^{2} \; ,
\label{5f}
\end{equation}
where
\begin{equation}
 h(z)=1+\frac{z^{2}}{1+ \sqrt{1+z^{2}}} \; .
\end{equation}
It is convenient to expand
\begin{equation}
{\rm ln}[1-{\rm e}^{-2n{\rm h}(z) (\alpha -1)}]=-\sum_{k=1}^
{\infty} {{\rm e}^{-2 n {\rm h}(z)(\alpha -1) k} \over k}.
\end{equation}
Inserting this expression into Eq.(\ref{xxx}) one can
compute first the sum over $n$, then the integral over $y$ and
finally the sum over $k$. The contribution of the $n=0$ term
in Eq.(\ref{xxx}) is negligible for $\alpha\sim 1$. The
final result is
\begin{equation}
E_{12}\sim E_{ex}\sim-\ell{\hbar c\pi^{3}\over 360 a^{2}}
\ {1\over (\alpha -1)^{3}}
+ O({1\over (\alpha -1)^2}) \; .
\end{equation}
As expected, in this
limit we obtain the proximity approximation \cite{derjaguin}
for the Casimir energy (see Eq.(\ref{ED})).

\section{Numerical results}\label{snum}

In this section we present the numerical results corresponding to
the semiclassical and exact calculations described in the previous
sections.

Before comparing the results, it is worth to note that
%%%the semiclassical approximation is not accurate for a single infinite
%%% cylinder. Indeed
, as described in Section III, the semiclassical e.m. energy for
an isolated cylinder vanishes (Eq. (\ref{semi1cil})), while the
exact energy for a cylinder of radius $a$ is $-0.01356 \; \ell \;
\hbar c/a^2$ (Eq.(\ref{exfin})). Therefore, in principle one could
compare the semiclassical Casimir energy with the full exact
energy given in Eq.(\ref{exfin}), or with the interaction energy
$E_{12}$. This leads to some ambiguities, although the second
possibility seems more appropriate because only trajectories
bouncing off between both cylinders give a non vanishing
contribution to the semiclassical energy.

\begin{figure}[ht]
\begin{center}
\includegraphics[width=10cm,height=15cm]{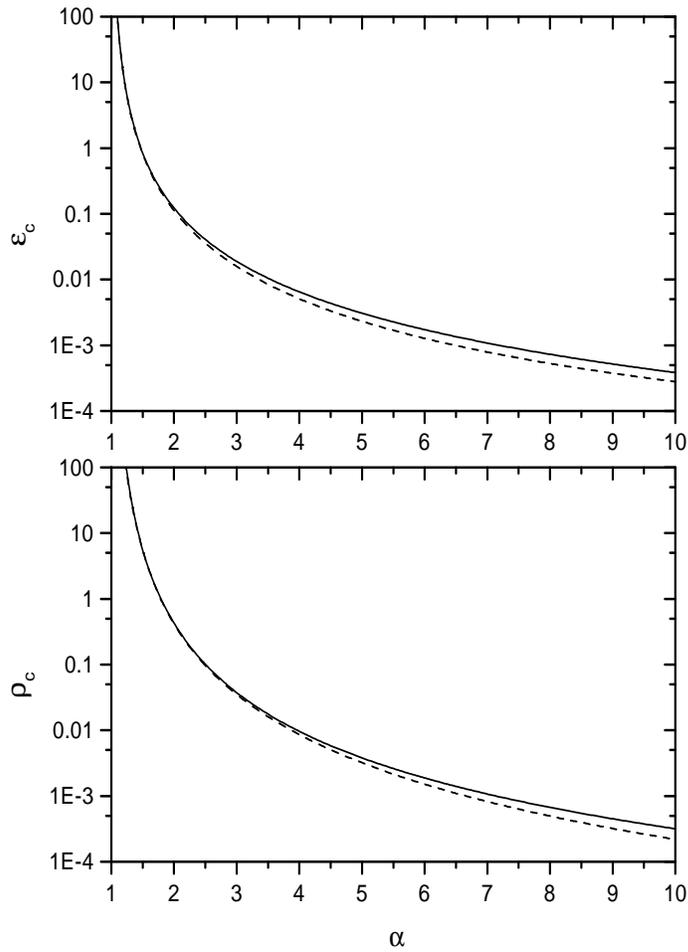}
\caption{Dimensionless Casimir interaction energy (upper panel) and pressure
(lower panel), defined in Eqs.(\ref{eadim}) and (\ref{adim}),
as a function of $\alpha = b/a$. In both panels the dashed line corresponds to the exact
result and the full line to the semiclassical result.} \label{fig4}
\end{center}
\end{figure}

In Fig. \ref{fig4} we display the exact results
(dashed line)
together with those corresponding to the semiclassical calculation
(full line) as a function of the ratio of the radii of the two
cylinders, $\alpha = b/a$.
For convenience we have defined the dimensionless
energies  as
\begin{equation} \label{eadim}
\epsilon_{12} = - \frac{a^2}{\hbar c}\ \frac{E_{12}}
{\ell} \;, \; \; \;  \;  \;  \epsilon^{sem} = - \frac{a^2}{\hbar c}\ \frac{E^{sem}}
{\ell} \; .
\end{equation}

As all energies are defined up to a constant, it is more 
meaningful to compare  the exact and semiclassical pressures on  
a given  cylinder. 
Specifically, we compare  the pressure on the inner cylinder due to the presence of the outer cylinder,
\begin{equation}
p_{12}=-\frac{1}{2\pi a\ell}\frac{\partial E_{12}}{\partial a} \; ,
\label{pres12}
\end{equation} 
with 
\begin{equation}
p^{sem} = - \frac{1}{2\pi a\ell}\frac{\partial E^{sem}}{\partial a} \; .
\label{presem}
\end{equation}
We also display in Fig. \ref{fig4} the exact and semiclassical
results for the dimensionless pressures
 \begin{equation}
\rho_{12} = \frac{2 \pi a^4}{\hbar c}\ p_{12} \;,  \; \; \; \; \; 
\rho^{sem} = \frac{2 \pi a^4}{\hbar c}\ p^{sem} \; .
\label{adim}
\end{equation}

As expected, the
discrepancies between the results increases with $\alpha$.
 One should remark that, as already
mentioned, the comparison between the exact and semiclassical
energies is subjected to some ambiguities that are not present in
the case of the pressures. Therefore, assuming  a typical
experimental precision of the order of 10 \%, such discrepancies
would be observable only for values  $\alpha > 4$ if the pressure
would be measured (see Fig.\ref{fig4} lower panel). Thus, it is
fair to conclude that, within a typical experimental error, the
semiclassical expressions lead to very good results for values of
$\alpha$ in the rather wide range $ 1 < \alpha  < 4$. It should be
stressed that, within that range, the Casimir interaction energies
and pressures decrease by several orders of magnitude.

The full
pressure on the inner cylinder is
\begin{equation}
p_{ex}=-\frac{1}{2\pi a\ell}\frac{\partial E_{ex}}{\partial a}
=p_{12}-0.01356{\hbar c\over \pi a^3} \; .
\label{presfull}
\end{equation}
For $\alpha\sim 1$ the attraction of the outer cylinder
dominates and the full pressure is positive. The
inner cylinder tends to expand. However,
when $\alpha$ is large, the self-pressure (second term in the
above equation) is bigger and the full pressure becomes negative.
The crossover takes place at $\alpha\sim 3.15$. It is remarkable
that the semiclassical approximation is accurate beyond
this critical value.

\begin{figure}[ht]
\begin{center}
\includegraphics[width=12cm,height=10cm]{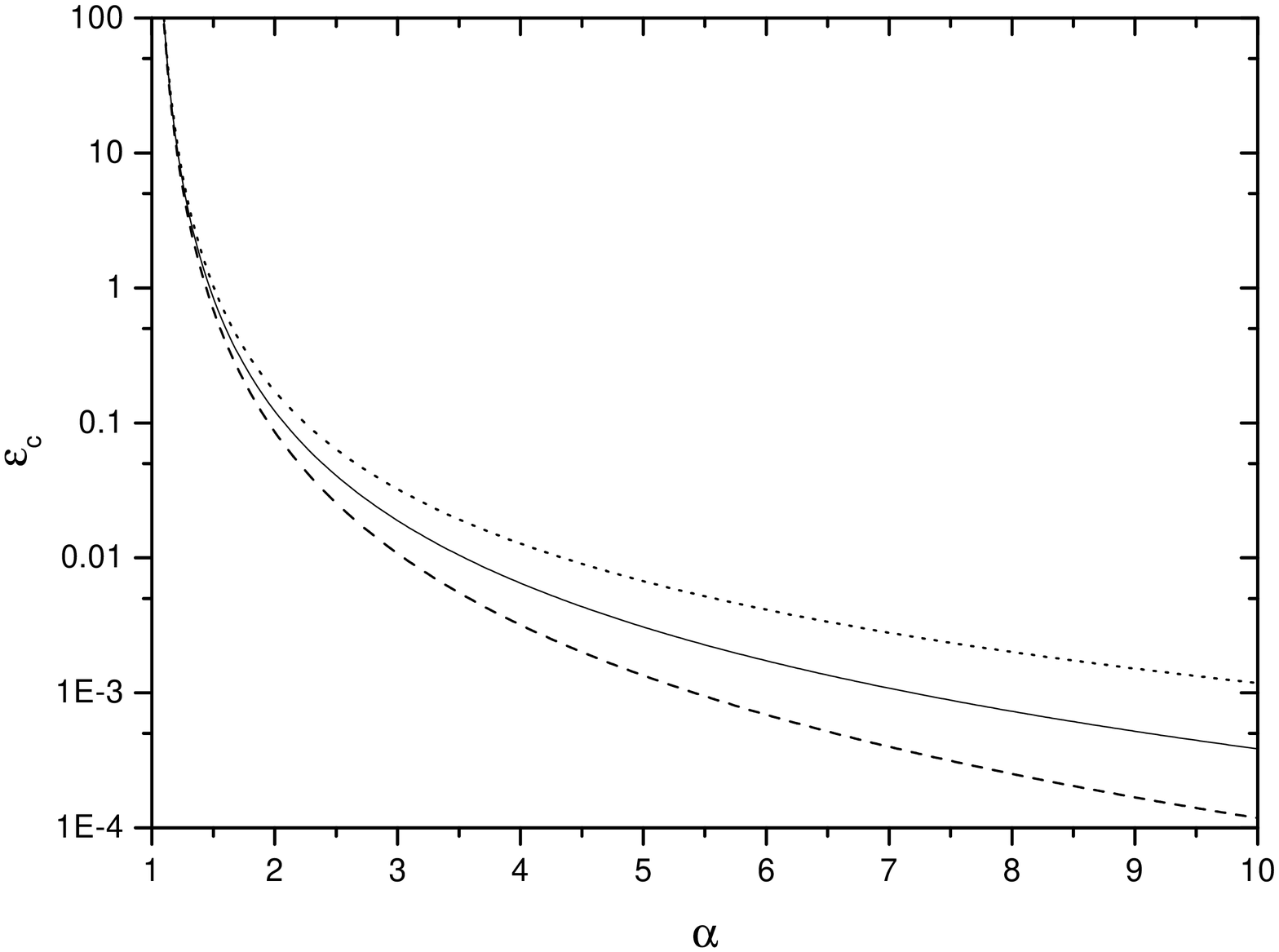}
\caption{Dimensionless Casimir interaction energy in the semiclassical
approximation (full line) as compared to the result obtained using
two naive ways of determining the relevant area which enters in the
proximity theorem: area of the inner cylinder (dashed
line); area of the outer cylinder (dotted line).} \label{fig5}
\end{center}
\end{figure}

Given the good results obtained within the semiclassical
approximation it is interesting to consider them in greater
detail. We have seen that in the case of the $w=0$ contribution
an explicit analytical expression can be obtained for arbitrary
values of $\alpha$. One might wonder how well
the exact results can be described by this expression.
Numerical calculation shows that if one would plot such contribution
in Fig. \ref{fig4} they would be completely indistinguishable
from those corresponding to the full semiclassical calculation.
In fact, even for $\alpha = 10$ the contributions with $w \ge 1$
represents less than 3\% of the total value of the energy or
pressure. Thus, it is clear that the semiclassical results
are completely dominated by the $w=0$ contribution which, in turn, is
dominated by the primitive self-retracing orbit $(v=1)$ and its first
repetitions $(v = 2, 3)$.
We have also mentioned that the expression corresponding to
the $w=0$ contribution, Eq. (\ref{ew0}), reduces exactly to that
given by the proximity theorem in the limit
$\alpha \rightarrow 1$. However, as explained at the end of
Sec.III, it is not clear how to extrapolate the use of
such theorem away from that limit. The basic problem is that
there is an ambiguity on which area should be considered for
the differential surfaces. In fact, one might either take
the area given by the radius of the inner cylinder, the one
given by the radius of the external cylinder or any combination
in between (e.g. average radius, geometric mean radius, etc).
The situation is illustrated in
Fig. \ref{fig5} where our semiclassical results (full line)
are compared with the extrapolation of the proximity theorem
using the two extreme alternatives: area given by the inner
cylinder (dashed line) and that given by the outer cylinder
(dotted line). Obviously, any other sensible choice would
lie between these two curves. Although for values of $\alpha$
very close to one this ambiguity is completely harmless, already
at such small values $\alpha \approx 1.12 $ it implies an uncertainty
of the order of 10 \%. The present semiclassical calculation removes completely such
ambiguity indicating that indeed the geometric mean radius
is the right choice.

\section{Conclusions}\label{scon}

In this article we have computed the Casimir interaction energy
for two infinite concentric cylinders exactly and using an
approximate semiclassical method. To perform the semiclassical
calculation, we have extended  the method introduced in
Ref.\cite{schaden2}, valid for non-symmetric configurations with
isolated periodic trajectories, to the case of symmetric
configurations in which families of non isolated periodic orbits
contribute significantly in the semiclassical limit. Technically,
this involves the use of the Balian-Bloch or Berry-Tabor trace
formulae instead of the Gutzwiller formula. As for the exact
calculation, we have used the mode-by-mode summation method of
Ref.\cite{piro}, but using a cutoff instead of zeta function
regularization.

The final result for the semiclassical Casimir energy (Eq.(\ref{main})),
is dominated by the self-retracing periodic orbit  $(w=0)$ and
its repetitions. Moreover,
it can be interpreted as a proximity approximation with an effective
area given by the geometric mean of the areas of both cylinders.
While this choice for  the area was previously derived for proximity forces
in non-symmetric configurations \cite{schaden1,blocki}, the semiclassical
calculation provides a justification for the case of an integrable
cavity.
We have found that, surprisingly, the semiclassical result describes
accurately the Casimir pressure beyond the  range of
validity of the proximity theorem. Indeed, while this theorem is
expected to be valid when $b-a\ll a$, the semiclassical result
reproduces the exact pressure within a 10 \% up to $b = 4 a$.

As a side point, we remark that the semiclassical approximation
fails to reproduce the electromagnetic Casimir energy for
 an isolated infinite cylinder.
Unlike the case of a rectangular parallelepiped \cite{schaden2}, the
 Dirichlet and Neumann contributions  have opposite signs.
Taking  into account that for cavities with axial symmetry, the
 electromagnetic Casimir energy is the sum of Dirichlet and Neumann
scalar contributions, we obtained a vanishing result for the
semiclassical energy (other approximations also give
a null result for the cylinder \cite {bd2,maclay}).
This shows that the semiclassical method may not
work for cavities with one of the dimensions much larger
than the others, as was suggested in Ref. \cite{schaden2}.

The semiclassical approach described here could be applied to
other integrable configurations of conductors, like two concentric spheres.
Moreover, it can  be extended to incorporate  effects that could be relevant
experimentally, like small surface deformations or roughness.
We are currently investigating these problems.

\section{ACKNOWLEDGMENTS}
This work  was
supported by Universidad de Buenos Aires, Conicet,
and Agencia Nacional de Promoci\'on Cient\'\i fica y Tecnol\'ogica, Argentina.

\appendix
\section{}

In this Appendix we describe with some detail the steps followed  to obtain 
the oscillating contributions $\rho^{osc}_{12,I}$
and  $\rho^{osc}_{12,II}$, Eqs. (\ref{roI}) and (\ref{roII}).

As we have mentioned in Sec.\ref{s2}
the oscillating contribution for  non-relativistic particles confined in a
bidimensional annular region  $\odot$, $\rho^{osc}_{\odot} (E_m)$, has been
previously derived (see e.g. Richter's book in Ref.\cite{brack}).
It can  been written as a sum of two terms, each one associated respectively
 to the contributions from type-I and type-II PO's. In Sec.\ref{s2}
we have characterized these PO, whose  lengths   are 
$L_{vw} = 2 \, v \; b \sin (\pi w /v) $ for   type-I PO, and
$\bar L_{vw} = 2 \, v \; b \;  \sqrt{ \left( 1+ \; \left(\frac{a}{b}\right)^2
\; - 2 \; \frac{a}{b}  \; \cos (\pi w /v) \right)}$  for type-II PO (see
the description that precedes Eqs. (\ref{lI}) and (\ref{lII})).
Taking this into account we can write
 $\rho^{osc}_{\odot} (E_m) \; = \rho^{osc}_{\odot,I} (E_m) +  
\rho^{osc}_{\odot,II} (E_m)$ with  $E_m = {\hbar \; k } ^2 / 2 \; m $ 
 \cite{brack} and

\begin{eqnarray}
\rho^{osc}_{\odot,I} (E_m) & = &  \sum_{w \ge 1} \;  \sum_{v \ge \hat v}
\sqrt{\frac{2}{\pi}} \; \frac{m}{ {\hbar}^2 \; k^{1/2}} \; 
\frac{L_{vw}^{3/2}}{v^2}  \; \cos\left(k \; L_{vw} \pm \frac{ v \; \pi}{2} + 
\frac{\pi}{4} \right) \; ,
\label{ronrI} \\
\rho^{osc}_{\odot,II} (E_m) & = &   \sum_{w \ge 0} \; \sum_{v \ge \hat v}  f_{vw}
\; 2 \sqrt{\frac{ 2 }{\pi}} \frac{b^2 \; m}{(\hbar)^2 \;(k \; {\bar L_{vw}})
^{1/2}} \; A_{vw} \;
\sin \left( k \;  \bar L_{vw}  + \; \frac{\pi}{4} \right)  \label{ronrII} \; .
\end{eqnarray}

 $A_{vw}$ has been defined in Eq.(\ref{avw}). Therefore  in order to  
obtain the oscillating contributions 
to the density of modes  for photons confined in the region  $\odot$  
we have  to  replace in  Eqs. (\ref{ronrI}) and 
(\ref{ronrII}), ${\hbar \; k } ^2 / 2 \; m \rightarrow p/c $ and 
$ (\hbar \; k) / m \rightarrow c$. 
This is trivial  and leads to  $\rho^{osc}_{\odot,I} (k)$ and 
$\rho^{osc}_{\odot,II}(k)$  with  $ k = E / \hbar c $.
 
To derive the oscillating contribution to the spectral density 
in the region between the two coaxial cylinders we have to perform the
integral Eq.(\ref{2d3d}), that we write here in a slightly different form,

\begin{equation}
\rho^{osc}_{12, po} (E = \hbar c \; k ) = \frac{\ell}{\hbar \; c \pi} \;
\int_{0}^{k} \frac {k}{\sqrt{k^2 - k_{z}^2}} \; \rho^{osc}_{\odot, po}(\sqrt{k^2 - k_{z}^2}) \; d k_z \;,
\end{equation}

to emphasize with the subscript  $po = I$ or $ II$ the contributions 
from the two types of PO's.
$ \rho^{osc}_{\odot, po} (\sqrt{k^2 - k_{z}^2})$ is obtained
after replacing $k \rightarrow \sqrt{k^2 - k_{z}^2}$ in 
 $\rho^{osc}_{\odot,I} (k )$ and in  
$\rho^{osc}_{\odot,II}(k)$.
The integrals are straightforward  to perform to leading order in $\hbar$, 
and as a result we obtain   Eqs.(\ref{roI}) and (\ref{roII}).

\pagebreak

\newpage
\begin{thebibliography}{99}

\bibitem{Casimir} H. B. G. Casimir, Proc. K. Ned. Akad. Wet. \textbf{51},
793 (1948); V.M. Mostepanenko and N.N. Trunov, {\it
The Casimir effect and its applications}, Clarendon, London (1997);
M. Bordag, {\it The Casimir effect 50 years later}, World Scientific,
Singapore (1999); P. Milonni, {\it The quantum vacuum}, Academic Press,
San Diego (1994); G. Plunien, B. Muller and W. Greiner, Phys.
Rep. {\bf 134}, 87 (1986).

\bibitem{roberto} G. Bressi, G. Carugno, R. Onofrio and G. Ruoso,
Phys. Rev. Lett. {\bf 88}, 041804 (2002).

\bibitem{Lamoreaux} S. K. Lamoreaux, Phys. Rev. Lett. {\bf 78}, 5 (1997).

\bibitem{Mohideen} U. Mohideen and A. Roy, Phys. Rev. Lett. {\bf 81}, 4549
(1998);
B.W. Harris, F. Chen and U. Mohideen, Phys. Rev. A {\bf 62}, 052109 (2000).

\bibitem{chan1} H.B. Chan {\it et al.}, Science {\bf 291}, 1941 (2001).

\bibitem{chan2} H.B. Chan {\it et al.}, Phys. Rev. Lett. {\bf 87},
1801 (2001).

\bibitem{Bordag} M. Bordag, U. Mohideen and V.M. Mostepanenko,
Phys. Rep. {\bf 353}, 1 (2001).

\bibitem{boyer} T.H. Boyer, Phys. Rev. {\bf 174}, 1764 (1968).

\bibitem{piro}V.V. Nesterenko and I.G. Pirozhenko, Phys. Rev. D{\bf 57},
1284 (1997).

\bibitem{bowers}M.E. Bowers and C.R. Hagen, Phys. Rev. D {\bf 59},
02007 (1999).

\bibitem{deraad}L.L. DeRaad Jr. and K. Milton, Ann. Phys. (N.Y.){\bf 136},
229 (1981).

\bibitem{nesterenko}K.A. Milton, A.V. Nesterenko and V.V. Nesterenko,
Phys. Rev. D {\bf 59}, 105009 (1999).

\bibitem{romeo}P. Gosdzinsky and A. Romeo, Phys. Lett. B {\bf 441},
265 (1998).

\bibitem{derjaguin}B.V. Derjaguin and I.I. Abriksova, Sov. Phys. JETP
{\bf 3}, 819 (1957); B.V. Derjaguin, Sci. Am. {\bf 203}, 47
(1960).

\bibitem{schaden1} M. Schaden and L. Spruch, Phys. Rev. Lett. {\bf 84},
459 (2000).

\bibitem{saha}A.A. Saharian, Phys. Rev. D{\bf 63}, 125007 (2001)

\bibitem{sahats}A.A. Saharian, ICTP preprint, IC/2000/14.


\bibitem{aclaracion} See Plunien et al, in
Ref[1].

\bibitem{bd2} R. Balian and B. Duplantier, {\sl Ann. Phys.}
{\bf 112}, 165 (1978)

\bibitem{schaden2} M. Schaden and L. Spruch, Phys. Rev. A {\bf 58},
935 (1998).

\bibitem{w} H. Weyl, Nach. Akad. Wiss. G\~ottingen 110, (1911).
\bibitem{bb1} R. Balian and C. Bloch,  {\sl Ann. Phys.} {\bf 60}, 401 (1970);
{\bf 63}, 592 (1971).
\bibitem{bal} H. P. Baltes and E. R. Hilf: {\sl Spectra of finite systems}
( B - I Wissenschaftsverlag, Mannheim, 1976).
\bibitem{bb2} R. Balian and C. Bloch,  {\sl Ann. Phys.} {\bf 69}, 76 (1972).
\bibitem{bt} M. V. Berry and M. Tabor, {\sl Proc. R. Soc. London, Ser. A.} {\bf 349}, 101 (1976).
\bibitem{gutz} M. C. Gutzwiller in
{\sl Chaos in Classical and Quantum Mechanics} (Springer-Verlag, New York,
1990).
\bibitem{brack} For a review see for example:
M. Brack and R. K. Bhaduri in {\sl Semiclassical
Physics} (Addison-Wesley Publishing Company, Masachussets, 1997);
K. Richter in {\sl Semiclassical Theory of Mesoscopic Quantum
Systems} (Springer, Berlin, 2000).
\bibitem{bd} R. Balian and B. Duplantier, {\sl Ann. Phys.} {\bf 104}, 300
(1977).
\bibitem{majo} M. J. S\'anchez, unpublished.
\bibitem{blocki}
J. Blocki, J. Randrup, W.J. Swiatecki and
F. Tsang, {\sl Ann. Phys.}{\bf 105}, 427 (1977).
\bibitem{aclaracion2} For simplicity, to compute the limit
$R,R_1,R_2\rightarrow\infty$ we assumed $R_2/R_1=\alpha$. The final
result for the Casimir energy does not depend on this
particular choice.
\bibitem{maclay} G.J. Maclay, H. Fearn, P. W. Milonni quant-ph/0105002



\end{thebibliography}
\end{document}